\documentclass[conference,letterpaper]{IEEEtran}
\usepackage{bm,cite}
\IEEEoverridecommandlockouts

\usepackage[table]{xcolor}
\usepackage{hhline}
\usepackage[acronym,shortcuts]{glossaries}
\usepackage{subfigure}
\usepackage{color}
\usepackage{graphicx}
\usepackage{amsmath}

\usepackage{amsmath,amsfonts}
\usepackage{physics, gensymb}
\usepackage{algorithm}
\usepackage{algpseudocode}
\usepackage{comment}
\usepackage{pgfplots} 

\newacronym[plural=SNRs,firstplural= signal-to-noise ratios (SNRs)]{SNR}{SNR}{signal-to-noise ratio}
\newacronym{OC-SVM}{OC-SVM}{one-class support vector machine}
\newacronym{OCC}{OCC}{one-class classification}
\newacronym{ML}{ML}{machine learning}
\newacronym{AoA}{AoA}{angle of arrival}
\newacronym{5G}{5G}{fifth-generation}
\newacronym{PLA}{PLA}{physical layer authentication}
\newacronym{RF}{RF}{radio frequency}
\newacronym{SVM}{SVM}{support vector machine}
\newacronym{UE}{UE}{user equipment}
\newacronym{BS}{BS}{base station}
\newacronym{LL}{LL}{log-likelihood}
\newacronym{FA}{FA}{false alarm}
\newacronym{MD}{MD}{missed detection}
\newacronym{AWGN}{AWGN}{additive white Gaussian noise}
\newacronym{RMSE}{RMSE}{root mean square error}
\newacronym{MIMO}{MIMO}{multiple-input multiple-output}

\begin{document}
\bstctlcite{IEEEexample:BSTcontrol}

\setlength\unitlength{1mm}

\newcommand{\insertfig}[3]{
\begin{figure}[htbp]\begin{center}\begin{picture}(120,90)
\put(0,-5){\includegraphics[width=12cm,height=9cm,clip=]{#1.eps}}\end{picture}\end{center}
\caption{#2}\label{#3}\end{figure}}

\newcommand{\insertxfig}[4]{
\begin{figure}[htbp]
\begin{center}
\leavevmode \centerline{\resizebox{#4\textwidth}{!}{\input
#1.pstex_t}}
%\vspace*{-0.2in}
\caption{#2} \label{#3}
\end{center}
\end{figure}}

\long\def\comment#1{}

% bb font symbols
% \DeclareMathOperator*{\argmax}{arg\,max}
% \DeclareMathOperator*{\argmin}{arg\,min}

\newfont{\bbb}{msbm10 scaled 700}
\newcommand{\CCC}{\mbox{\bbb C}}
\newcommand{\ssn}[1]{{\scriptscriptstyle{{#1}}}}
\newcommand{\ssr}[1]{{\scriptscriptstyle{\mathrm{#1}}}}
\newcommand{\ssnb}[1]{{\scriptscriptstyle{({#1})}}}
\newcommand{\ssrb}[1]{{\scriptscriptstyle{(\mathrm{#1})}}}

% replace `c' with `!\vthickline'
% replace \hline with \hthickline
\newcommand{\vthickline}{\vrule width 0.8pt}
\newcommand{\hthickline}{\noalign{\hrule height 0.80pt}}

\newfont{\bb}{msbm10 scaled 1100}
\newcommand{\CC}{\mbox{\bb C}}
\newcommand{\PP}{\mbox{\bb P}}
\newcommand{\RR}{\mbox{\bb R}}
\newcommand{\QQ}{\mbox{\bb Q}}
\newcommand{\ZZ}{\mbox{\bb Z}}
\newcommand{\FF}{\mbox{\bb F}}
\newcommand{\GG}{\mbox{\bb G}}
\newcommand{\EE}{\mbox{\bb E}}
\newcommand{\NN}{\mbox{\bb N}}
\newcommand{\KK}{\mbox{\bb K}}

% Vectors

\newcommand{\hermit}{\mathsf{H}}

\newcommand{\av}{{\bf a}}
\newcommand{\bv}{{\bf b}}
\newcommand{\cv}{{\bf c}}
\newcommand{\fv}{{\bf f}}
\newcommand{\gv}{{\bf g}}
\newcommand{\hv}{{\bf h}}
\newcommand{\iv}{{\bf i}}
\newcommand{\jv}{{\bf j}}
\newcommand{\kv}{{\bf k}}
\newcommand{\lv}{{\bf l}}
\newcommand{\mv}{{\bf m}}
\newcommand{\nv}{{\bf n}}
\newcommand{\ov}{{\bf o}}
\newcommand{\qv}{{\bf q}}
\newcommand{\rv}{{\bf r}}
\newcommand{\sv}{{\bf s}}
\newcommand{\tv}{{\bf t}}
\newcommand{\uv}{{\bf u}}
\newcommand{\wv}{{\bf w}}
\newcommand{\vv}{{\bf v}}
\newcommand{\xv}{{\bf x}}
\newcommand{\yv}{{\bf y}}
\newcommand{\zv}{{\bf z}}
\newcommand{\zerov}{{\bf 0}}
\newcommand{\onev}{{\bf 1}}
\newcommand{\avr}{\av_\text{R}}

\newcommand{\varphivd}{\varphiv_{\text{D}}}
\newcommand{\varphiva}{\varphiv_{\text{A}}}
\newcommand{\phia}{\phi_{\text{A}}}
\newcommand{\thetaa}{\theta_{\text{A}}}
\newcommand{\phid}{\phi_{\text{D}}}
\newcommand{\thetad}{\theta_{\text{D}}}

\newcommand{\phis}{\phi_{\text{S}}}
\newcommand{\varphivs}{\varphiv_{\text{S}}}
\newcommand{\thetas}{\theta_{\text{S}}}
\newcommand{\avy}{\av_y}
\newcommand{\avz}{\av_z}

\newcommand{\avry}{\av_\text{R,y}}
\newcommand{\avrz}{\av_\text{R,z}}

\def\u{{\bf u}}

% Matrices

\newcommand{\Am}{{\bf A}}
\newcommand{\Bm}{{\bf B}}
\newcommand{\Cm}{{\bf C}}
\newcommand{\Dm}{{\bf D}}
\newcommand{\Em}{{\bf E}}
\newcommand{\Fm}{{\bf F}}
\newcommand{\Gm}{{\bf G}}
\newcommand{\Hm}{{\bf H}}
\newcommand{\Id}{{\bf I}}
\newcommand{\Jm}{{\bf J}}
\newcommand{\Km}{{\bf K}}
\newcommand{\Lm}{{\bf L}}
\newcommand{\Mm}{{\bf M}}
\newcommand{\Nm}{{\bf N}}
\newcommand{\Om}{{\bf O}}
\newcommand{\Pm}{{\bf P}}
\newcommand{\Qm}{{\bf Q}}
\newcommand{\Rm}{{\bf R}}
\newcommand{\Sm}{{\bf S}}
\newcommand{\Tm}{{\bf T}}
\newcommand{\Um}{{\bf U}}
\newcommand{\Wm}{{\bf W}}
\newcommand{\Vm}{{\bf V}}
\newcommand{\Xm}{{\bf X}}
\newcommand{\Ym}{{\bf Y}}
\newcommand{\Zm}{{\bf Z}}
\newcommand{\Onem}{{\bf 1}}
\newcommand{\Zerom}{{\bf 0}}

% Calligraphic

%\newcommand{\Ac}{{\cal A}}
\newcommand{\Bc}{{\cal B}}
\newcommand{\Cc}{{\cal C}}
\newcommand{\Dc}{{\cal D}}
\newcommand{\Ec}{{\cal E}}
\newcommand{\Fc}{{\cal F}}
\newcommand{\Gc}{{\cal G}}
\newcommand{\Hc}{{\cal H}}
\newcommand{\Ic}{{\cal I}}
\newcommand{\Jc}{{\cal J}}
\newcommand{\Kc}{{\cal K}}
\newcommand{\Lc}{{\cal L}}
\newcommand{\Mc}{{\cal M}}
\newcommand{\Nc}{{\cal N}}
\newcommand{\Oc}{{\cal O}}
\newcommand{\Pc}{{\cal P}}
\newcommand{\Qc}{{\cal Q}}
\newcommand{\Rc}{{\cal R}}
\newcommand{\Sc}{{\cal S}}
\newcommand{\Tc}{{\cal T}}
\newcommand{\Uc}{{\cal U}}
\newcommand{\Wc}{{\cal W}}
\newcommand{\Vc}{{\cal V}}
\newcommand{\Xc}{{\cal X}}
\newcommand{\Yc}{{\cal Y}}
\newcommand{\Zc}{{\cal Z}}

% Bold greek letters

\newcommand{\alphav}{\hbox{\boldmath$\alpha$}}
\newcommand{\betav}{\hbox{\boldmath$\beta$}}
\newcommand{\gammav}{\hbox{\boldmath$\gamma$}}
\newcommand{\deltav}{\hbox{\boldmath$\delta$}}
\newcommand{\etav}{\hbox{\boldmath$\eta$}}
\newcommand{\lambdav}{\hbox{\boldmath$\lambda$}}
\newcommand{\kappav}{\hbox{\boldmath$\kappa$}}
\newcommand{\epsilonv}{\hbox{\boldmath$\epsilon$}}
\newcommand{\nuv}{\hbox{\boldmath$\nu$}}
\newcommand{\muv}{\hbox{\boldmath$\mu$}}
\newcommand{\zetav}{\hbox{\boldmath$\zeta$}}
\newcommand{\phiv}{\hbox{\boldmath$\phi$}}
\newcommand{\psiv}{\hbox{\boldmath$\psi$}}
\newcommand{\thetav}{\hbox{$\boldsymbol\theta$}}
\newcommand{\tauv}{\hbox{\boldmath$\tau$}}
\newcommand{\omegav}{\hbox{\boldmath$\omega$}}
\newcommand{\xiv}{\hbox{\boldmath$\xi$}}
\newcommand{\sigmav}{\hbox{\boldmath$\sigma$}}
\newcommand{\piv}{\hbox{\boldmath$\pi$}}
\newcommand{\rhov}{\hbox{\boldmath$\rho$}}

\newcommand{\boldone}{{ {\boldsymbol{1}} }}

\newcommand{\varthetav}{\hbox{\boldmath$\vartheta$}}
\newcommand{\varphiv}{\hbox{\boldmath$\varphi$}}
\newcommand{\vpv}{\boldsymbol{\varphi}}
\newcommand{\vtv}{\boldsymbol{\vartheta}}

\newcommand{\Gammam}{\hbox{\boldmath$\Gamma$}}
\newcommand{\Lambdam}{\hbox{\boldmath$\Lambda$}}
\newcommand{\Deltam}{\hbox{\boldmath$\Delta$}}
\newcommand{\Sigmam}{\hbox{\boldmath$\Sigma$}}
\newcommand{\Phim}{\hbox{\boldmath$\Phi$}}
\newcommand{\Pim}{\hbox{\boldmath$\Pi$}}
\newcommand{\Psim}{\hbox{\boldmath$\Psi$}}
\newcommand{\psim}{\hbox{\boldmath$\psi$}}
\newcommand{\chim}{\hbox{\boldmath$\chi$}}
\newcommand{\omegam}{\hbox{\boldmath$\omega$}}
\newcommand{\Thetam}{\hbox{\boldmath$\Theta$}}
\newcommand{\Omegam}{\hbox{\boldmath$\Omega$}}
\newcommand{\Xim}{\hbox{\boldmath$\Xi$}}

% mixed symbols

\newcommand{\sinc}{{\hbox{sinc}}}
\newcommand{\diag}{{\hbox{diag}}}
\renewcommand{\det}{{\hbox{det}}}
\newcommand{\sign}{{\hbox{sign}}}
\renewcommand{\arg}{{\hbox{arg}}}
\newcommand{\cov}{{\hbox{cov}}}
\newcommand{\SINR}{{\sf sinr}}
\newcommand{\SNR}{{\sf snr}}
\newcommand{\Ei}{{\rm E}_{\rm i}}
\newcommand{\eqdef}{\stackrel{\Delta}{=}}
\newcommand{\defines}{{\,\,\stackrel{\scriptscriptstyle \bigtriangleup}{=}\,\,}}
\newcommand{\<}{\left\langle}
\renewcommand{\>}{\right\rangle}
\newcommand{\herm}{{\sf H}}
\newcommand{\trasp}{{\sf T}}
\renewcommand{\vec}{{\rm vec}}
\newcommand{\transp}{{\sf T}}
\newcommand{\calL}{\mbox{${\mathcal L}$}}
\newcommand{\calO}{\mbox{${\mathcal O}$}}

\newcommand{\Abc}{\mbox{$\boldsymbol{\mathcal{A}}$}}
\newcommand{\Bbc}{\mbox{$\boldsymbol{\mathcal{B}}$}}
\newcommand{\Cfd}{\mbox{$\boldsymbol{\mathcal{C}}$}}
\newcommand{\Dfd}{\mbox{$\boldsymbol{\mathcal{D}}$}}
\newcommand{\Efd}{\mbox{$\boldsymbol{\mathcal{E}}$}}
\newcommand{\Ffd}{\mbox{$\boldsymbol{\mathcal{F}}$}}
\newcommand{\Gfd}{\mbox{$\boldsymbol{\mathcal{G}}$}}
\newcommand{\Hbc}{\mbox{$\boldsymbol{\mathcal{H}}$}}
\newcommand{\Ifd}{\mbox{$\boldsymbol{\mathcal{I}}$}}
\newcommand{\Jfd}{\mbox{$\boldsymbol{\mathcal{J}}$}}
\newcommand{\Kfd}{\mbox{$\boldsymbol{\mathcal{K}}$}}
\newcommand{\Lfd}{\mbox{$\boldsymbol{\mathcal{L}}$}}
\newcommand{\Mfd}{\mbox{$\boldsymbol{\mathcal{M}}$}}
\newcommand{\Nfd}{\mbox{$\boldsymbol{\mathcal{N}}$}}
\newcommand{\Ofd}{\mbox{$\boldsymbol{\mathcal{O}}$}}
\newcommand{\Pfd}{\mbox{$\boldsymbol{\mathcal{P}}$}}
\newcommand{\Qfd}{\mbox{$\boldsymbol{\mathcal{Q}}$}}
\newcommand{\Rfd}{\mbox{$\boldsymbol{\mathcal{R}}$}}
\newcommand{\Sfd}{\mbox{$\boldsymbol{\mathcal{S}}$}}
\newcommand{\Tfd}{\mbox{$\boldsymbol{\mathcal{T}}$}}
\newcommand{\Ufd}{\mbox{$\boldsymbol{\mathcal{U}}$}}
\newcommand{\Vfd}{\mbox{$\boldsymbol{\mathcal{V}}$}}
\newcommand{\Wfd}{\mbox{$\boldsymbol{\mathcal{W}}$}}
\newcommand{\Xfd}{\mbox{$\boldsymbol{\mathcal{X}}$}}
\newcommand{\Yfd}{\mbox{$\boldsymbol{\mathcal{Y}}$}}
\newcommand{\Zfd}{\mbox{$\boldsymbol{\mathcal{Z}}$}}

% \algnewcommand{\Inputs}[1]{%
%   \State \textbf{Inputs:}
% }
% \algnewcommand{\Initialize}[1]{%
%   \State \textbf{Initialize:}
%   \Statex \hspace*{\algorithmicindent}\parbox[t]{.8\linewidth}{\raggedright #1}
%\title{Impersonation Attacks on AoA-Based Authentication in Analog Arrays \marco{AoA-Based Physical Layer Authentication in Analog Array MIMO systems under Impersonation Attacks}}
\title{AoA-Based Physical Layer Authentication in Analog Arrays under Impersonation Attacks}


\author{Muralikrishnan Srinivasan\IEEEauthorrefmark{1}, Linda Senigagliesi\IEEEauthorrefmark{2}, Hui Chen\IEEEauthorrefmark{3}, Arsenia Chorti\IEEEauthorrefmark{4}, Marco Baldi\IEEEauthorrefmark{2}, Henk Wymeersch\IEEEauthorrefmark{3}\\
\IEEEauthorrefmark{1}Indian Institute of Technology (BHU), Varanasi, India \IEEEauthorrefmark{2}Università Politecnica delle Marche, Italy\\
\IEEEauthorrefmark{3}Chalmers University of Technology, Sweden, \IEEEauthorrefmark{4}ETIS UMR 8051 / CY Paris University, ENSEA, CNRS, France\thanks{This work was supported by the Swedish Research Council (VR grant 2023-03821), by Hexa-X-II, part of the European Union’s Horizon Europe research and innovation programme under Grant Agreement No 101095759 and by project SERICS (PE00000014) under the MUR National Recovery and Resilience Plan funded by the European Union - NextGenerationEU.}
}


\maketitle

\begin{abstract}
%    With the increasing reliance on wireless localization, e.g., in navigation and location-based services, its trustworthiness arises as an important research topic. Features derived from channel state information such as the received signal strength (RSS), \ac{AoA}, angle of departure (AoD), etc., have been proposed as authentication factors in recent works. %As these systems heavily depend on the physical-layer properties of exchanged messages, including Message Received Signal Strength, Time of Arrival, and Angle of Arrival (AoA), securing the integrity of location results demands a comprehensive analysis encompassing potential physical-layer manipulations by attackers. 
    We discuss the use of \ac{AoA} as an authentication measure in analog array \ac{MIMO} systems. A base station equipped with an analog array authenticates users based on the \ac{AoA} estimated from certified pilot transmissions, while active attackers manipulate their transmitted signals to mount impersonation attacks. We study several attacks of increasing intensity (captured through the availability of side information at the attackers) and assess the performance of \ac{AoA}-based authentication using one-class classifiers. Our results show that some attack techniques with knowledge of the combiners at the verifier are effective in falsifying the AoA and compromising the security of the considered type of physical layer authentication.
\end{abstract}

\glsresetall

\section{Introduction}
%In the fifth generation (5G) wireless systems, two authentication protocols have been introduced, namely the 5G authentication and key agreement protocol -- for devices equipped with subscriber identification modules (SIM) cards -- and the extensible authentication protocol for authentication and key agreement (EAP-AKA), for SIM-less devices. Despite enhancements introduced with respect to long-term evolution (LTE) authentication, there still remain significant open issues, e.g., i) sidelink authentication outside the coverage of a base station; ii) certificate-less authentication in massive connectivity regimes; iii) fast authentication for ultra low latency applications, to name but a few. In this framework, 
\Ac{PLA} is gaining momentum in the realm of wireless communication systems due to its ability to be deployed relatively easily in device-to-device setups without the need for a cumbersome public key infrastructure \cite{xie2020survey}. {Unlike conventional cryptographic methods, PLA authenticates devices or users based on unique signal characteristics observed at the physical layer. %By acquiring and building a database based on wireless channel properties like signal strength, time of arrival, \ac{AoA}, channel response, etc., PLA verifies the communicating parties' authenticity.
In addition to its fast processing and high interoperability in heterogeneous systems, such as its ability to work jointly with or complement upper-layer authentication, \ac{PLA} enables the construction of multi-factor authentication schemes for enhanced security\cite{xie2020survey, mitev_fast_2020}.
%Additionally, PLA alleviates the requirement for certificates and is known for its rapid process \cite{xie2020survey, mitev_fast_2020}. %Firstly, it provides information-theoretic security by introducing uncertainty to adversaries, making it computationally unbreakable. Secondly, \ac{PLA} enables quick differentiation between legitimate and rogue transmitters without upper-layer processing, saving computational complexity. Lastly, 
%Furthermore, \ac{PLA} can offer high interoperability in heterogeneous systems and can be used jointly or complement upper-layer authentication, allowing the construction of multi-factor authentication schemes for enhanced security.
A typical PLA scenario is shown in Fig. \ref{fig:Authscheme}: the training process involves acquisition of initial features from transmissions of a legitimate user (Alice), associating them with her identity by leveraging some manual process or higher-layer identification protocol. Subsequent transmissions from Alice are then recognized by comparison with the signals acquired during training, and possibly distinguished from those of an attacker (Eve) attempting to impersonate Alice by manipulating transmitted signals through precoding.

% In a typical PLA scenario shown in Fig. \ref{fig:Authscheme} (a) and (b), the process involves acquiring legitimate transmitter characteristics and associating them with its identity. During subsequent transmissions, the legitimate transmitter is authenticated by comparing the signals acquired during training, leaving illegitimate transmitters unauthenticated.} 

\begin{figure}[th]
    \centering
    \subfigure[]{\includegraphics[width=0.4\textwidth]{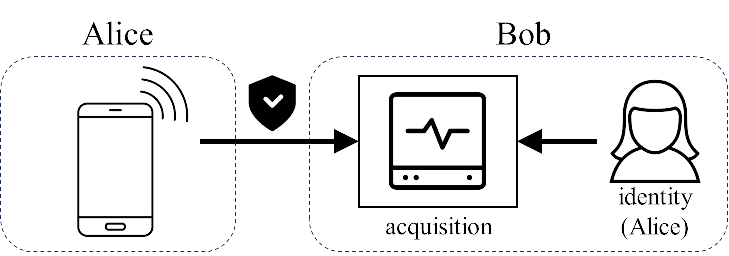}} 
    \subfigure[]{\includegraphics[width=0.4\textwidth]{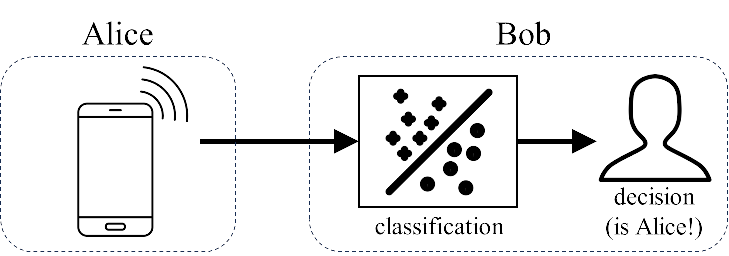}} 
    \subfigure[]{\includegraphics[width=0.4\textwidth]{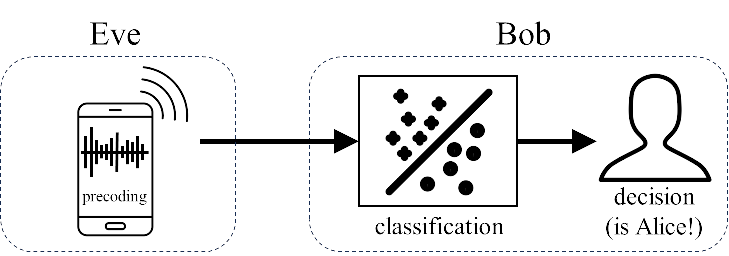}}
    \caption{Considered physical layer authentication scenario: (a) \emph{Offline Training phase:} Alice's transmission is guaranteed to be authentic, the corresponding signal is acquired by Bob and associated with her identity (b) \emph{Online classification phase:} Alice is recognized by the classifier through comparison of the newly acquired signal with her original signal, collected during training (c) \emph{Authentication system under successful impersonation attack:} Eve transmits a precoded signal with the aim to confuse Bob's classifier and make it recognize the received signal as being transmitted by Alice.}\vspace{-7mm}
    \label{fig:Authscheme}
\end{figure}

\ac{PLA} can be used either in challenge-response authentication protocols or in tag-based authentication protocols \cite{gil2023}, and includes hardware-based and channel-based authentication. As an example of challenge-response \ac{PLA}, hardware fingerprints, referred to as physical unclonable functions (PUFs), serve as unclonable unique device identifiers \cite{hou2014physical, sankhe2019oracle, mitev2023}. Recently, channel controllability using reflective intelligent surfaces (RIS) has also been considered in challenge-response protocols \cite{tomasin2022}. 
%{The fundamental concept of tag-based authentication involves creating an \emph{authentication tag} based on channel characteristics.} 
Channel-based \ac{PLA} employs various channel features for authentication \cite{xiao2007fingerprints, wang2016wireless, maurer2000authentication, lai2009authentication, senigagliesi2020comparison, srinivasan2023smart}. In some recent works on channel-based \ac{PLA}, like \cite{abdelaziz2019enhanced, xu2022physical, casari2022physical}, the use of the \ac{AoA} as a unique identity feature has been proposed, and also used to identify Sybil attacks in robotic networks \cite{gil2017guaranteeing}.  
Significantly, there has been a surge of interest in integrating \ac{ML} into PLA \cite{casari2022physical, fang2018learning,senigagliesi2020comparison}, %For instance, in \cite{senigagliesi2020comparison}, authors introduced an \ac{ML}-driven PLA approach leveraging mmWave \ac{MIMO} channels, incorporating azimuth and elevation \ac{AoA}s, along with carrier frequency offset (CFO). The work explores various classification methods based on \ac{ML}, including nearest neighbor (NN) and \ac{SVM} algorithms, examining both their binary and one-class versions.
as well as model-based approaches for location spoofing \cite{li2023channel,li2024optimized}.

Similar to any authentication scheme, \ac{AoA}-based \ac{PLA} solutions are susceptible to active attacks, in particular to impersonation/spoofing. Such attacks aim to deceive the verifier (Bob) -- possibly by employing suitable precoders -- to mis-classify Eve as Alice due to induced errors in the estimated \ac{AoA}. Robustness against such attacks has been explored only in a handful of works up to now, like \cite{abdelaziz2016security, pham2023machine}. The former focuses on jamming attacks and discusses the optimality of maximum likelihood-based \ac{AoA} estimation. The latter focuses on spoofing attacks on \ac{AoA} estimation in \ac{MIMO} systems with digital arrays at the verifier. 
%In these circumstances, it was shown that i) such attacks are possible under the form of injection attacks as long as the attacker possesses one additional antenna compared to the legitimate users, and ii) it is possible to alleviate such attacks by the use of pilot randomization techniques. 
%
In digital arrays, each receive antenna corresponds to a dedicated \ac{RF} chain and the verifier estimates the \ac{AoA} by examining phase variation across the array. 
%does not extend to \ac{MIMO} systems employing analog arrays. In more detail, in the case of digital arrays each receive antenna corresponds to a dedicated \ac{RF} chain and the verifier estimates the \ac{AoA} by examining phase variation. However, 
In the case of \emph{analog arrays} with a single \ac{RF} chain, the \ac{AoA} is estimated through multiple pilot transmissions, probing different angles with appropriate beamforming vectors for each transmission. 
In view of the fact that low-cost IoT devices can possibly not afford digital array transceivers, the study of spoofing attacks in \ac{AoA}-based \ac{PLA} systems using analog arrays becomes very important and motivates the current study.

In this paper, we demonstrate that analog arrays are much more vulnerable to spoofing attacks. Our contributions are as follows: (i)
%\begin{itemize}    
 %   \item 
 We consider a standard authentication protocol in a novel context, where a verifier, in this case the \ac{BS}, is equipped with an analog array and identifies a node, exploiting a one-class classifier trained using the estimated \acp{AoA} of the legitimate node; (ii)
 %   \item  
 We investigate impersonation attacks by a malicious node and study the impact of the impersonation attacks on the estimated \acp{AoA}; (iii)
 %   \item 
 We study the impact of the impersonation attacks on the ML-based classifier's performance.
%\end{itemize}

\subsubsection*{Notation} Vectors $\bm{x}$ are denoted in bold, transpose as $\bm{x}^\top$, Hermitian as $\bm{x}^{\mathsf{H}}$, and complex conjugate as  $\bm{x}^*$. 
\section{Authentication Model}\label{sec:auth}
In this section, we first describe the system model, followed by an  authentication protocol, focusing on the \ac{AoA} as the primary feature. 

\subsection{System Model}
Consider Alice's single-antenna\footnote{We have opted for a simple yet non-trivial system model to gain fundamental insights. This approach allows us to establish foundational principles and understand core mechanisms without unnecessary complexities.}
 \ac{UE} located at coordinates $\bm{x}^{\text{A}}=[{x}^{\text{A}}_1, {x}^{\text{A}}_2]^\top$. %Alice represents a legitimate node, whose communications to the \ac{BS} are authorized. 
The \ac{BS}, corresponding to Bob, is situated at the origin $[0, 0]^\top$ and is equipped with an analog array, i.e., $N$ receive antennas and a single \ac{RF} chain, subject to the constraint $N>1$. 
Alice transmits a sequence of $T>1$ uplink pilot signals $\bm s^{\text{A}} = [s^{\text{A}}_1, \ldots, s^{\text{A}}_T]^\top$ to Bob satisfying the constraint $\Vert\bm s^{\text{A}}\Vert^2= 1$, where $T$ is the number of pilot transmissions. The received signal at Bob can be expressed as 
\begin{align}
y_t = \sqrt{P} h^{\text{A}} \bm{w}^{\mathsf{H}}_t \bm{a}(\theta^{\text{A}}) s^{\text{A}}_t + n_t, \quad t=1,2,\ldots, T, \label{eq:Observation_nominal}
\end{align}
where $P$ represents the transmitted power from Alice, $|h^{\text{A}}|=\lambda /(4 \pi d^{\text{A}})$  is the channel amplitude between Alice and Bob (as a function of the distance $d^{\text{A}} = \Vert\bm x^{\text{A}}\Vert$ and the wavelength $\lambda$),  $n_t$ is complex \ac{AWGN} with variance $\sigma^2/2$ per real dimension, $\bm{w}_t$ is the beamforming vector at time $t$, and $\bm{a}(\theta^{\text{A}})$ stands for the array steering vector\footnote{For a linear antenna array, the $n$-th element of the array steering vector is given by $[\bm{a}(\theta^{\text{A}})]_n = \exp(\jmath \pi n \sin(\theta^{\text{A}}))$, for $n=1,2,\ldots, N$} determined by the angle-of-arrival (\ac{AoA}) $\theta^{\text{A}}$, where $\theta^{\text{A}}=\arctan({x}^{\text{A}}_1/{x}^{\text{A}}_2)$. 
The beamforming vectors 
$\bm{w}_t$ are configured to directional beams, i.e., $\bm{w}_t=\bm{a}(\theta_t)$, where $\theta_t$ denotes the probing direction between $[-90 \degree, 90 \degree]$. Also, {$\sigma^2 = N_0 W$, where $N_0$ is the noise power spectral density 
and $W$ is the bandwidth.}
%Here, $P$ represents the transmitted power from Alice, $P_{r}^{\text{A}}$ represents the received power from Alice at Bob, 
%$d^{\text{A}} = \norm{\bm x^{\text{A}}}$ represents the distance of Alice from Bob, $\lambda$ represents the wavelength, $\bm{w}_t \in \mathbb{C}^{N\times 1}$ is the combining vector for the $t$-th transmission, $\bm{a}(\theta^{\text{A}})$ stands for the array steering vector determined by the angle-of-arrival (\ac{AoA}) $\theta^{\text{A}}$, where $\theta^{\text{A}}=\arctan({x}^{\text{A}}_1/{x}^{\text{A}}_2)$.  \footnote{Since $\bm{w}^{\mathsf{H}}_t \bm{a}(\theta^{\text{A}})$ results in only real values, real-valued channel gain is considered}. 
%For the case of analog arrays, estimating the \ac{AoA} ($\theta^{\text{A}}$) requires at least $T\geq 2$ pilot transmissions. 
It is assumed that Bob possesses knowledge of $\bm s^{\text{A}}$, based on which, the \ac{AoA} $\hat \theta$ at the BS can be estimated. %, for instance, by setting $s^{\text{A}}_t=1$ for all $t$. 
% Based on this shared knowledge, Bob estimates the \ac{AoA} at the BS,   given by $\hat \theta$.

%\hui{why such as, \ac{AoA} is our only feature right? do we have distance? Ersi: we might need the distance to distinguish users on the same \ac{AoA}} 

\subsection{Physical Layer Authentication Protocol}

We consider a standard authentication protocol in which Bob needs to identify Alice, based on the estimated \ac{AoA}. Authentication protocols include an offline training phase and an online verification phase, which are schematically depicted in Fig.~\ref{fig:Authscheme}-(a) and Fig.~\ref{fig:Authscheme}-(b), and described next.
\begin{itemize}
    \item 
\emph{Offline Training Phase:}
 features (in our case estimated AoAs $\hat \theta$) are recorded by Bob for each legitimate user; a corresponding database of user/node identifiers (IDs) and features is then created. 
%In other words, we assume that in the offline phase, the authenticator is allowed to estimate the AoAs and map them correctly to the identities of all users / nodes that may subsequently request to be authenticated.
 %We report the training phase of a user, Alice, by Bob. During this phase, 
 The transmissions of Alice received by Bob are guaranteed to be authentic by higher-layer protocols. 
From an authentication perspective, this corresponds to a \ac{OCC} scenario, where only the positive class (target class) is present during training, while the negative classes (non-target classes) remain unknown \cite{senigagliesi2020comparison}.

%OCC relies on two crucial parameters: the distance $d(x)$ between the sample to classify and the target class, and the threshold $\theta_d$ on the value of the distance. Formally, a new instance is classified as belonging to the positive class if its distance from the target class is below the threshold:
%\begin{equation}
%f_{\text{oc}}(x) = \mathbb{I}(d(x) < \theta_d)
%\label{eq:dec_fun}
%\end{equation}
%where $\mathbb{I}( \cdot )$ is an indicator function, and $f_{\text{oc}}(x)$ is the decision function—a binary function expressing acceptance of the object $x$ into the target class. Various metrics can be employed to evaluate $d(x),$ with the Euclidean distance being a common choice.  Drawing on prior knowledge, the receiver assigns labels to new instances during an on-line phase, determining which instances to authenticate and which ones to reject.
% \begin{figure}
%     \centering
% \includegraphics[width=0.85\linewidth]{Figures/onboarding.eps}
%     \caption{Offline onboarding phase: Alice and Charlie transmissions are guaranteed to be authentic, and the corresponding signals are acquired by Bob and associated with their identities.}
%     \label{fig:onboarding}
% \end{figure}

\item \emph{Online Verification Phase:}
Alice begins by announcing her identity to Bob and transmitting pre-agreed pilot sequences. This allows Bob to estimate Alice's \ac{AoA} and authenticate her using an \ac{OCC} classifier trained in the previous phase.\footnote{In precise terms, we define \textit{identification} as a process enabling the authenticator to recognize the node's identity without explicit inquiry. On the other hand, \textit{authentication} refers to a procedure where the node initially declares its identity and subsequently provides proof to verify that they are indeed the claimed identity.} 
\end{itemize}

%\subsection{Problem Formulation}
The purpose of this work is to study the robustness of such a standard authentication protocol to several new impersonation attacks, tailored to \ac{AoA} estimation with analog arrays.

\section{Impersonation Attacks}

Any authentication system is subject to attacks, notably impersonation/spoofing attacks. %These aim to trick Bob into mis-classifying transmissions originating from an attacker, referred to as Eve, a single-antenna \ac{UE}. 
We assume that the attacker (Eve) is provided with a single-antenna UE and employs suitable precoding with the aim that the received signals prompt Bob's classifier to mis-classify them as originating from Alice, as schematically illustrated in Fig.~\ref{fig:Authscheme}-(c). 

\subsection{General Attack Model}
Eve, located at coordinates $\bm{x}^{\text{E}}=[{x}^{\text{E}}_1, {x}^{\text{E}}_2]^\top$, induces an array steering vector  $\bm{a}(\theta^{\text{E}})$ at Bob, determined by the \ac{AoA} $\theta^{\text{E}}$, where $\theta^{\text{E}}=\arctan({x}^{\text{E}}_1/{x}^{\text{E}}_2)$. Eve manipulates the \ac{AoA}-based authentication of Alice by altering the transmitted signal $\bm s^{\text{E}} =[s^{\text{E}}_1, \ldots, s^{\text{E}}_T]^\top$ to Bob, while satisfying the constraint $\Vert\bm s^{\text{E}}\Vert^2= 1$. %, where $T$ is the number of pilot transmissions.
The received signal at Bob originating from Eve can be expressed as:
\begin{align}
{y}_t = \sqrt{P} h^{\text{E}} \bm{w}^{\mathsf{H}}_t \bm{a}(\theta^{\text{E}}) s^{\text{E}}_t + {n}_t, \quad t=1,2,\ldots, T,
\end{align}
where $P$ is Eve's transmission power (here for simplicity set equal to the transmission power of Alice), $|h^{\text{E}}|=\lambda/(4 \pi d^{\text{E}})$ is the channel amplitude, in which $d^{\text{E}} = \Vert\bm x^{\text{E}}\Vert$ is the distance of Eve from Bob, 
%Here, $P_{r}^{\text{E}}$ represents the received power at Bob from Eve, $h^{\text{E}}$ represents the channel gain between Eve and Bob, $\bm{a}(\theta^{\text{E}})$ stands for the array steering vector determined by the \ac{AoA} $\theta^{\text{E}}$, where 
$\theta^{\text{E}}=\arctan({x}^{\text{E}}_1/{x}^{\text{E}}_2)$, and ${n}_t$ denotes the \ac{AWGN} component  with variance $\sigma^2/2$ per real dimension. %The received power from Eve at Bob is given by $P_{r}^{\text{E}} = P \lambda^2/(4 \pi d^{\text{E}})^2$, where $d^{\text{A}} = \norm{\bm x^{\text{E}}}$ is the distance of Eve from Bob. 
Eve can manipulate the transmitted signal or its statistical properties in different ways, depending on the knowledge she has regarding Alice. 
In the following we describe some strategies that Eve might follow to manipulate her signals and the corresponding assumptions on her knowledge about Alice.

\subsection{Random Attack}
% \noindent 
%\emph{Attacker Knowledge:} 
Eve knows when to transmit and the duration of the transmission. 
% \noindent 
%\emph{Attack Operation:} 
Eve generates $s^{\text{E}}_t=\exp(\jmath \phi_t)/\sqrt{T}$, where $\phi_t \sim \mathcal{U}[0,2\pi]$. 
% \noindent 
%\emph{Attack Impact:} 
Since $s^{\text{E}}_t\neq s^{\text{A}}_t$, the \ac{AoA} estimation capability at Bob is compromised. However, it is unlikely that Eve can impersonate Alice with this attack. 

\subsection{Code-based attack}
%\noindent \emph{Attacker Knowledge:} 
Eve knows when to transmit, and the duration of the transmission.  Eve also knows the combiners $\bm{w}_t$ (which can be interpreted as a code, hence the name code-based attack), the pilot $\bm s^{\text{A}}$,   and  Alice's \ac{AoA} $\theta^{\text{A}}$. Eve does not know  $\theta^{\text{E}}$. 
%\noindent \emph{Attack Operation:}
%The code-based attack essentially follows a similar procedure as the location-based attack, with a distinction in the pilot manipulation stage. 
 Here, Eve can manipulate $s^{\text{E}}_t$ as
\begin{align}
     s^{\text{E}}_t  = \alpha^{\text{E}} {\bm{w}^{\mathsf{H}}_t \bm{a}(\theta^{\text{A}})} s^{\text{A}}_t,
\end{align}
where $\alpha^{\text{E}}$ is a normalization value set to meet the constraint $\Vert\bm s^{\text{E}}\Vert^2= 1$, i.e., $\alpha^{\text{E}}=({{\sum_{t=1}^{T}|\bm{w}^{\mathsf{H}}_t \bm{a}(\theta^{\text{A}})s^{\text{A}}_t|^2}})^{-1/2}$. %.\footnote{In particular $\alpha^{\text{E}}=\left({\sqrt{\sum_{t=1}^{T}|\bm{w}^{\mathsf{H}}_t \bm{a}(\theta^{\text{A}})s^{\text{A}}_t|^2}}\right)^{-1}$.} 
%using the following expression:
%\begin{equation}
 %   s^{\text{E}}_t = \frac{\beta_t}{\norm{\bm \beta}} \text{for}  \: t=1,2, \ldots, T
%\end{equation}
%where $\bm \beta = \left[ \beta_1, \beta_2, \ldots, %\beta_T \right]$ and 
 %   \begin{equation}
  %  \beta_t=\bm{w}^{\mathsf{H}}_t \bm{a}(\theta^{\text{A}})\: \text{for}  \: t=1,2, \ldots, T.
   % \end{equation} 
Subsequently, the received signal at time $t$ at the \ac{BS} is given by 
\begin{align}
   {y}_t & = \sqrt{P} h^{\text{E}} \alpha^{\text{E}} \bm{w}^{\mathsf{H}}_t \bm{a}(\theta^{\text{E}})  {\bm{w}^{\mathsf{H}}_t \bm{a}(\theta^{\text{A}})} s^{\text{A}}_t +n_t.
\end{align}
%  \begin{equation}
%        {y}_t =\sqrt{P_{r}^{\text{E}}} \bm{w}^{\mathsf{H}}_t \bm{a}(\theta^{\text{E}}) %\frac{\bm{w}^{\mathsf{H}}_t \bm{a}(\theta^{\text{A}})}{\norm{\bm \beta}} + n_t \quad t=1,2,\ldots, T,
%    \end{equation}
%\noindent \emph{Attack Impact:} 
This manipulation causes the \ac{BS} to perceive a signal arriving from \emph{both} the impersonator \ac{AoA} $\theta^{\text{E}}$ and the true \ac{AoA} $\theta^{\text{A}}$, providing opportunities for Eve to impersonate Alice. 

 \subsection{Location-based attack} 
 %\noindent \emph{Attacker Knowledge:}
 In this case, Eve acquires knowledge of the combiners $\bm{w}_t$,  the pilot $\bm s^{\text{A}}$,  the target angle $\theta^{\text{A}}$ and her own angle $\theta^{\text{E}}$, leveraging information about both the \ac{BS} and her own location. 
 %\noindent \emph{Attack Operation:}
 Eve can manipulate $s^{\text{E}}_t$ as follows:
\begin{align}
     s^{\text{E}}_t  = \alpha^{\text{E}} \frac{\bm{w}^{\mathsf{H}}_t \bm{a}( \theta^{\text{A}})(\bm{w}^{\mathsf{H}}_t \bm{a}(\theta^{\text{E}}))^*}{|\bm{w}^{\mathsf{H}}_t \bm{a}(\theta^{\text{E}})|^2}s^{\text{A}}_t,
\end{align}
 where $\alpha^{\text{E}}$ is again a normalization value set to meet the constraint $\Vert\bm s^{\text{E}}\Vert^2= 1$.
%\begin{equation}
%    s^{\text{E}}_t = \frac{\beta_t}{\norm{\bm \beta}} \:\text{for}  \: t=1,2, \ldots, T
%\end{equation}
%    where $(.)^*$ represents the conjugate operation. Here,
 %   \begin{equation}
  %      \beta_t=\frac{\bm{w}^{\mathsf{H}}_t \bm{a}( \theta^{\text{A}})(\bm{w}^{\mathsf{H}}_t \bm{a}(\theta^{\text{E}}))^*}{|\bm{w}^{\mathsf{H}}_t \bm{a}(\theta^{\text{E}})|^2}, \quad t=1,2,\ldots, T.
   % \end{equation}
    %The manipulated signal, $s^{\text{E}}_t$, is then transmitted. 
    The received signal at Bob at time $t$ is given by
    \begin{align}
        y_t &  =\sqrt{P} h^{\text{E}} \alpha^{\text{E}} \bm{w}^{\mathsf{H}}_t \bm{a}(\theta^{\text{E}})  \frac{\bm{w}^{\mathsf{H}}_t \bm{a}( \theta^{\text{A}})(\bm{w}^{\mathsf{H}}_t \bm{a}(\theta^{\text{E}}))^*}{|\bm{w}^{\mathsf{H}}_t \bm{a}(\theta^{\text{E}})|^2}s^{\text{A}}_t +n_t\\
        & = \sqrt{P} h^{\text{E}} \alpha^{\text{E}}   \bm{w}^{\mathsf{H}}_t \bm{a}( \theta^{\text{A}}) s^{\text{A}}_t +n_t.
    \end{align}
  %  \begin{equation}
   %     y_t  = \sqrt{P_{r}^{\text{E}}}  \frac{\bm{w}^{\mathsf{H}}_t \bm{a}(\theta^{\text{A}})}{\norm{\bm \beta}} + n_t, \quad t=1,2,\ldots, T.
   % \end{equation}
   %\noindent \emph{Attack Impact:} 
This manipulation causes the \ac{BS} to perceive only a signal arriving from $\theta^{\text{A}}$, effectively eliminating any trace of the true angle $\theta^{\text{E}}$. However, the effectiveness of the attack depends on the power of $\alpha^{\text{E}}$, which Eve cannot control. In particular, if a combiner $\bm{w}_t$ is such that $|\bm{w}^{\mathsf{H}}_t \bm{a}(\theta^{\text{E}})| \ll |\bm{w}^{\mathsf{H}}_t \bm{a}( \theta^{\text{A}})|$, 
%Eve must dedicate a lot of her power on $s^{\text{E}}_t $, 
the overall potency of the attack is reduced.

\section{Numerical Results}

In this section, we assess the impact of impersonation attacks on the considered authentication protocol. 
%the estimated AoA and impact on the classifier, presenting the attack outcomes in terms of the \ac{LL} function and the classifier outcomes in terms of classification accuracy, probability of \ac{FA}, and \ac{MD}. 

\subsection{Authentication Protocol and Performance Metrics}
The authentication protocol comprises two stages: the \ac{AoA} estimator and the \ac{OCC}, generating the decisions. 

\subsubsection{\ac{AoA} Estimation}
We adopt maximum likelihood estimation of $\theta^{\text{A}}$   from \eqref{eq:Observation_nominal} in the presence of an unknown complex channel gain $h^{\text{A}}$, leading to 
\begin{align} \label{eq:ML}
    \hat{\theta}=\arg \min_{\theta} \Vert \bm{y}- \hat{h}(\theta) \bm z (\theta) \Vert^2,
\end{align}
%\marco{
%Maybe it could be written like this, for better clarity:
%\begin{align}
 %   \hat{\theta}=\arg \min_{\theta} \Vert \bm{y}- \hat{h}(\theta) \bm z (\theta) \Vert^2,
%\end{align}
%}
where $\bm{y}= [y_{1}, y_{2}, \ldots, y_{T}]^\top$, $[\bm z]_t = \bm w_t^{\mathsf{H}} \bm a(\theta)$, and $\hat{h}(\theta) = {\bm {z}^{\mathsf{H}}(\theta) \bm y}/{\norm{\bm{z}(\theta)}^2}$.

\subsubsection{Classifier}

For authentication, we employ \ac{OC-SVM} as the classifier at Bob \cite{senigagliesi2020comparison}. \ac{OC-SVM} aims to encapsulate the majority of training data within a hypersphere $R = \{\bm x \in \mathbb{R}^N | f_{\text{oc}}(\bm x) > 0 \}$, where $\bm x$ is the feature vector and $f_{\text{oc}}(\bm x)$ is the decision function. New samples are accepted or rejected based on the decision function $f_{\text{oc}}(\bm x)$: if $f_{\text{oc}}(\bm x) > 0$, the message is accepted; otherwise, it is rejected.
To assess the effectiveness of this decision process, we exploit the classical metrics based on the probability of \ac{FA} and \ac{MD}.
A \ac{FA} occurs when Bob wrongly rejects a message from Alice, while an \ac{MD} happens when a message from Eve is mistakenly accepted by Bob as authentic. %These metrics are evaluated using the Confusion Matrix \cite{Stehman1997}, which provides a comprehensive summary of the classification process outcomes. 
The probability of \ac{FA} is calculated as:
\begin{equation}
P_{\textsc{FA}} = \frac{\textsc{FN}}{\textsc{TP}+\textsc{FN}},
\label{eq:pfa}
\end{equation}
where $\textsc{FN}$ and $\textsc{TP}$ represent the number of false negatives and true positives, respectively. Similarly, the probability of MD is computed as:
\begin{equation}
    \label{eq:pmd}
    P_{\textsc{MD}} = \frac{\textsc{FP}}{\textsc{FP}+\textsc{TN}},
\end{equation}
where $\textsc{FP}$ and $\textsc{TN}$ represent the number of false positives and true negatives, respectively. Finally, overall accuracy is defined as:
\begin{equation}
    \label{eq:acc}
    \text{Acc} = \frac{\textsc{TP}+ \textsc{TN}}{\textsc{TP}+\textsc{TN} + \textsc{FP}+\textsc{FN}}.
\end{equation}

%To illustrate impersonation attacks, To estimate \ac{AoA} $\hat \theta$, the \ac{BS} uses \ac{LL} systematically scans beams in a grid spanning angles $\theta_k$ from $-90 \degree$ to $90  \degree$, with $k=1,2,\ldots K$ and a resolution of $K = 1000$ angles. For each angle $\theta_k$, corresponding to different AoAs, the $\text{LL}(\theta_k)$ given by 
%\begin{equation}
 %   \text{LL}(\theta_k) =  \norm {\bm{y}- \hat \alpha \bm z }^2
%\end{equation}
%is recorded. Here, $\bm{y}= [y_{1}, y_{2}, \ldots, y_{T}]$ and $[\bm z]_t = \bm w_t^H \bm a(\theta_k)$ for $t=1,2, \dots, T$ and $\hat \alpha = \frac{\bm z^H \bm y}{\norm{\bm z}^2}$. The estimated \ac{AoA} $\hat \theta$ is recorded from the peak of the \ac{LL} function using the expression:
%\begin{equation}\label{eq:LL}
%    \hat \theta = \argmax_{\theta_k} {\text{LL}(\theta_k)}.
%\end{equation}

%\begin{table}[]
%    \centering
 %   \begin{tabular}{|c|c|}
  %  \hline
   %     Parameter & Value \\
    %    \hline
     %   \hline
      %  $N$ & $16$  \\
%        \hline
 %       $T$ & $17$  \\
  %      \hline
   %     $d^{\text{A}}$ & $10~$m\\
    %    \hline
     %   $\theta^{\text{A}}$ & $0\degree$\\
  %      \hline 
  %      Carrier Frequency & $2.5~$GHz\\
  %      \hline
  %     $W$ & $20~$MHz\\
  %      \hline
  %      P & $10~$dB\\
  %      \hline
  %%      $N_0$ & $-174$~dBm/Hz\\
   %     \hline
   % \end{tabular}
   % \caption{Simulation parameters}
   % \label{tab:simulation}%
%\end{table}

\subsection{Simulation Parameters}

We consider a system operating at 2.5 GHz with $W=20~\text{MHz}$ bandwidth and a transmit power $P=10~\text{dBm}$. The noise power spectral density is $N_0=-174$~dBm/Hz. Bob is equipped with $N=16$ antennas and expects $T=17$ transmissions. The beamforming vectors $\bm{w}_t$ are set to directional beams, denoted as $\bm{w}_t=\bm{a}(\theta_t)$, where $\theta_t$ represents the $T$ probing directions, uniformly spanning from $-90 \degree$ to $90 \degree$. Alice is located at $d^{\text{A}}=10~\text{m}$ with  $\theta^{\text{A}}=0\degree$, while Eve will have a variable location.

\definecolor{mycolor2}{rgb}{0.85000,0.32500,0.09800}%
\definecolor{mycolor3}{rgb}{1,1,1}%
\definecolor{mycolor4}{rgb}{0.92900,0.69400,0.12500}%
\definecolor{mycolor5}{rgb}{0.46600,0.67400,0.18800}%
\newcommand{\AliceRef}{\tikz[baseline]{\draw[blue,dashed,line width = 2.0pt](0,0.8mm) -- (5.3mm,0.8mm)}}
\newcommand{\BobRef}{\tikz[baseline]{\draw[mycolor2,dotted,line width = 2.0pt](0,0.8mm) -- (5.3mm,0.8mm)}}
\newcommand{\RandomAttack}{\tikz[baseline]{\draw[mycolor5,line width = 2.0pt](0,0.8mm) -- (5.3mm,0.8mm)}}
\newcommand{\CodeAttack}{\tikz[baseline]{\draw[mycolor4,line width = 2.0pt](0,0.8mm) -- (5.3mm,0.8mm)}}
\newcommand{\LocAttack}{\tikz[baseline]{\draw[black,line width = 2.0pt](0,0.8mm) -- (5.3mm,0.8mm)}}

\subsection{Results and Discussion}
\subsubsection{Impact of the Attacks on the Estimated \ac{AoA}}
To illustrate the effects of different attacks on the estimated \ac{AoA}, we position Eve at a distance of $10$ meters from the \ac{BS} with an \ac{AoA} of $\theta^{\text{E}} = 45 \degree$, and we plot the negative log-likelihood cost $\Vert \bm{y}- \hat{h}(\theta) \bm z (\theta) \Vert^2$ from \eqref{eq:ML} in Fig.~\ref{fig:LL}.  % As discussed in Section II. A, . At the \ac{BS}, we establish a grid of $1000$ evenly spaced angles $\theta$ within the range of $[-90 \degree, 90 \degree]$ to compute $[\bm z]_t$, $\hat{h}(\theta)$, and consequently the logarithm of the likelihood function expressed in Equation (\ref{eq:ML}). 
%\marco{For maximum clarity, I would suggest that we explicitly write down the formula by which we calculate the log-likelihood of the objective function, and also refer to it in the caption of the figure.}
The cost originating from Alice (\protect\AliceRef) has a distinct minimum at $\theta=0\degree$. Similarly, if Eve does not deploy any attack, the corresponding cost  (\protect\BobRef) has a minimum at $\theta=45\degree$. Under the random attack  (\protect\RandomAttack), this minimum is attenuated and shifted but does not generate a minimum around $\theta^{\text{A}}$, and hence  Eve is incapable of impersonating Alice. Nonetheless, a sufficient number of random attacks could potentially facilitate successful Denial of Service Attacks, undermining Bob's capability to estimate the \ac{AoA}. The code-based attack (\protect\CodeAttack) does somewhat better and an additional minimum is created around $0$ degree apart from the one at $45 \degree$. Both minima are rather shallow and shifted from their nominal value. In this case, Eve has opportunities to impersonate Alice. Finally, with the location-based attack (\protect\LocAttack) a new minimum appears at $\theta=0 \degree$, while the original minimum at $\theta= 45\degree$ disappears. The minimum is pronounced and sharp, indicating that this attack can be highly effective at impersonating Alice. 

%Fig. \ref{fig:LL} illustrates the \ac{LL}, where the true \ac{AoA} is $\theta^{\text{A}} = 0$, $\theta^{\text{E}} = 45 \degree$. Without an attack, a clear peak is observed at the true \ac{AoA} $45 \degree$. With a code-based attack, an additional peak is created $0$ degree apart from the one at $45 \degree$. With a location-based attack, a new peak appears at $\theta^{\text{A}}=45 \degree$, while the original peak at $\theta^{\text{E}} = 0\degree$ disappears. In contrast, a random attack fails to generate a peak at $0\degree$, rendering Eve incapable of impersonating Alice. Nonetheless, it's worth noting that a sufficient number of random attacks could potentially facilitate successful Denial of Service Attacks, undermining Bob's capability to estimate the \ac{AoA}.
\begin{figure}%[!th]
    \centering
    \input{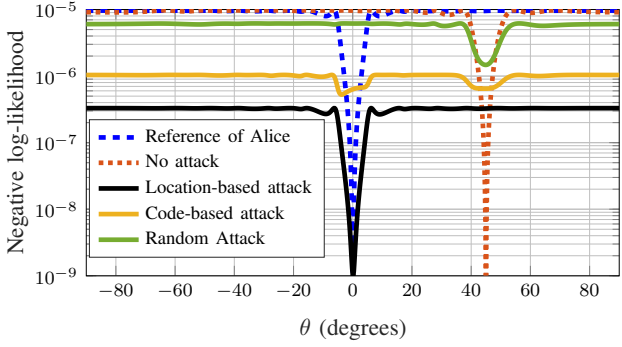}
    \vspace{-4mm}
    \caption{Negative log-likelihood cost function \eqref{eq:ML} as a function of the \ac{AoA}  for different attacks for $d^{\text E}= 10~$m. }\vspace{-5mm}
    \label{fig:LL}
\end{figure}

From now on, we will disregard the random attack and perform a more in-depth statistical analysis of the code-based and location-based attacks. To this end, we compute the \ac{RMSE}, defined as $(\mathbb{E}\{ (\hat \theta - \theta^{\text A})^2\})^{1/2}$ based on 1000 Monte Carlo trials for various $d^{\text{E}}$ and  $\theta^{\text{E}}$. 
%The likelihood function is calculated and the \ac{AoA} of Eve $\hat \theta_k$ is estimated using (\ref{eq:ML}) for $k=1,\ldots, 1000$ trials and \ac{RMSE} is calculated as $\text{RMSE} = \sqrt{1/1000\sum_{k=1}^{1000} \left( \hat \theta_k - \theta^{\text A}\right)^2}$. 
The \ac{RMSE} resulting from the code-based and location-based attacks is depicted in Fig. \ref{fig:estimatedAoA}. For the location-based attack, we observe that as the distance $d^{\text{E}}$ increases, the \ac{RMSE} increases due to a decrease in the \ac{SNR}, leading to the failure of the attack. Additionally, the \ac{AoA} estimation trend shows a predominantly monotonic behavior as $\theta^{\text{E}}$ increases but is also significantly influenced by nulls in the beam response. When the null of Eve's beam response precisely aligns with the main lobe of Alice's beam response, for example, at $\theta^{\text{E}} = 30 \degree$, achieving accurate emulation of Alice's response theoretically necessitates an infinite amount of power, resulting in $\alpha^{\text{E}}\to 0$.
Nevertheless, in a broader context, the overall trend suggests that with an increase in $\theta^{\text{E}}$, there is a corresponding increase in the \ac{RMSE} of the estimated \ac{AoA}. Notably, the \ac{RMSE} is more pronounced in the code-based attack compared to the location-based attack. This increased error is attributed to the presence of dual minima in the negative log-likelihood function, {leading to bias in the estimates, as discussed in the context of Fig.~\ref{fig:LL}. The code-based attack thus introduces bias, resulting in increased \ac{RMSE}; at angles exceeding $80 \degree$, a higher variance but reduced bias in \ac{AoA} estimates occurs for larger distances, leading to an overall smaller \ac{RMSE} compared to smaller distances.}

%Therefore, to ensure a practical and feasible scenario, it becomes essential to align the codebook at Bob such that the beams are not aligned with the nulls of Eve. 
%This constraint limits the maximum power that Eve can allocate to her transmitted signal, preventing unbounded power consumption and making the attack strategy more realistic within the constraints of the communication system.

\begin{figure}
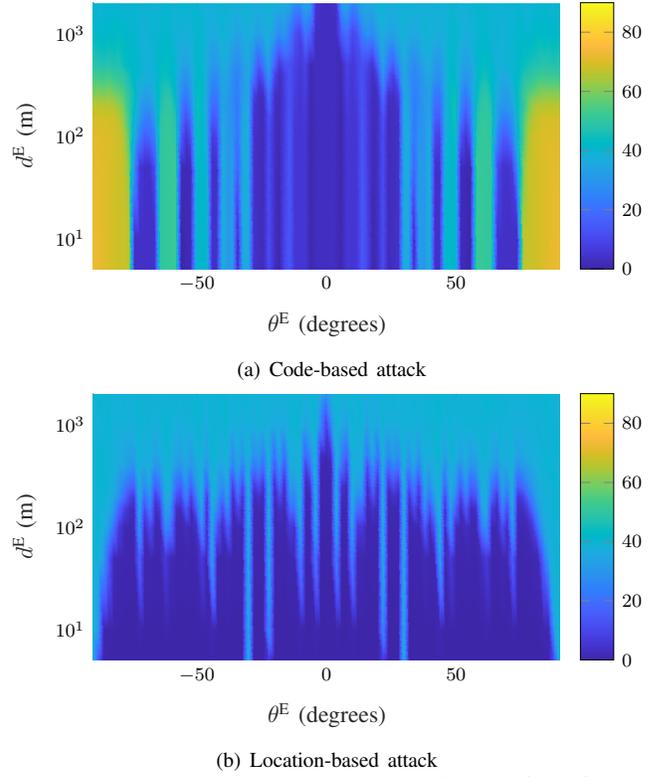

    \centering

    \subfigure[Code-based attack]{\input
    {figures_new/code_attack}
    %\caption{aa}
    } 
    \subfigure[Location-based attack]{\input{figures_new/location_attack}}
    \vspace{-3mm}
    \caption{RMSE in estimated angle in degrees vs distance of Eve from Bob and various \ac{AoA} of Eve.}\vspace{-2mm}
    \label{fig:estimatedAoA}
\end{figure}

\subsubsection{Impact of the Attacks on the Classifier}

% \begin{figure}
%     \centering
%     \includegraphics[width=\linewidth]{figures_new/Angle_Estimation.eps}
%     \caption{Heatmap showing the magnitude of error in the \ac{AoA} estimation by Bob from Eve's signal for various distances and angle of Eve. 
%     \\
%     \hui{I feel a figure with curves is better. Also, I think distance is a bit misleading as we do not use it; maybe change to SNR or transmit power?}}
%     \label{fig:securityULA_AoAestimation}
% \end{figure}
 The \ac{OC-SVM} classifier was trained on 1,000 \ac{AoA} samples from Alice, considering a distance between Alice and Bob equal to 10 meters. The test set consisted of 200,000 samples evenly split between Alice and Eve, comprising 100,000 legitimate signals and 100,000 attack signals. The reported results are averaged over 100 randomly selected training and test sets. Fig.~\ref{fig:acc_comparison} compares the accuracy achieved by the OC-SVM under both code-based and location-based attacks.
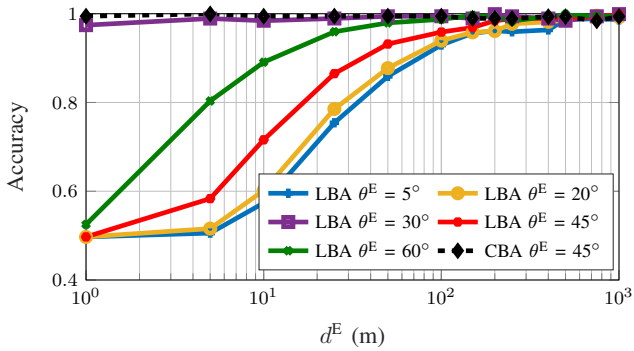
\begin{figure}%[ht]
\centering
\definecolor{mycolor1}{rgb}{0.00000,0.44700,0.74100}%
\definecolor{mycolor2}{rgb}{0.92900,0.69400,0.12500}%
\definecolor{mycolor3}{rgb}{0.49400,0.18400,0.55600}%
\definecolor{mycolor4}{rgb}{0.00000,0.49804,0.00000}%
\begin{tikzpicture}[scale=1\columnwidth/10cm,font=\footnotesize]
\begin{axis}[%
width=8 cm,
height=4 cm,
at={(0.932in,0.642in)},
scale only axis,
xmin=1,
xmax=1000,
xlabel style={font=\color{white!15!black}},
xlabel={$d^{\text E}$ (m)},
ymin=0.4    ,
ymax=1,
xmode=log,
ylabel style={yshift=-0.2cm},
yminorticks=true,
ylabel style={font=\color{white!15!black}},
ylabel={Accuracy},
axis background/.style={fill=white},
xmajorgrids,
xminorgrids,
ymajorgrids,
yminorgrids,
legend style={legend cell align=left, align=left,legend columns=2, fill opacity=0.8, draw=white!15!black, at={(0.99,0.4)}}
]

\addplot [color=mycolor1, line width=2.0pt, mark=+, mark options={solid, mycolor1}]
  table[row sep=crcr]{%
1	0.49699824120603\\
5	0.505656633165829\\
10	0.573579497487437\\
25	0.754206633165829\\
50	0.858873216080402\\
100	0.929519346733669\\
150	0.956029497487437\\
200	0.961489346733668\\
250	0.959329849246231\\
400	0.963544020100503\\
500	0.986735728643216\\
750	0.988740804020101\\
1000	0.986384271356784\\
2000	0.993675628140703\\
};
\addlegendentry{$\text{LBA }\theta{}^\text{E}\text{ = 5}^\circ$}

\addplot [color=mycolor2, line width=2.0pt, mark=o, mark options={solid, mycolor2}]
  table[row sep=crcr]{%
1	0.496967336683417\\
5	0.515865025125628\\
10	0.601536733668342\\
25	0.785336582914573\\
50	0.877422060301507\\
100	0.939385326633166\\
150	0.958023266331659\\
200	0.961081557788945\\
250	0.977115226130654\\
400	0.981645427135678\\
500	0.985456331658292\\
750	0.995548341708543\\
1000	0.992369648241206\\
2000	0.983985879396985\\
};
\addlegendentry{$\text{LBA }\theta{}^\text{E}\text{ = 20}^\circ$}

\addplot [color=mycolor3, line width=2.0pt, mark=square, mark options={solid, mycolor3}]
  table[row sep=crcr]{%
1	0.974263266331658\\
5	0.989137688442211\\
10	0.984208542713568\\
25	0.989232462311558\\
50	0.994157638190955\\
100	0.994314924623116\\
150	0.989204170854271\\
200	0.999015979899497\\
250	0.994137587939699\\
400	0.989162663316583\\
500	0.98423487437186\\
750	0.994161959798995\\
1000	0.999094824120603\\
2000	0.994335979899498\\
};
\addlegendentry{$\text{LBA }\theta{}^\text{E}\text{ = 30}^\circ$}

\addplot [color=red, line width=2.0pt, mark=asterisk, mark options={solid, red}]
  table[row sep=crcr]{%
1	0.497089296482412\\
5	0.583415879396985\\
10	0.715945226130653\\
25	0.865008291457286\\
50	0.931723567839196\\
100	0.95869959798995\\
150	0.968863567839196\\
200	0.984313718592965\\
250	0.988473316582914\\
400	0.990032211055276\\
500	0.991775376884422\\
750	0.993111155778895\\
1000	0.998693316582915\\
2000	0.998941256281407\\
};
\addlegendentry{$\text{LBA }\theta{}^\text{E}\text{ = 45}^\circ$}

\addplot [color=mycolor4, line width=2.0pt, mark=x, mark options={solid, mycolor4}]
  table[row sep=crcr]{%
1	0.525102613065327\\
5	0.803360954773869\\
10	0.890751708542714\\
25	0.959383015075377\\
50	0.979358643216081\\
100	0.987401055276382\\
150	0.996205879396985\\
200	0.992898844221105\\
250	0.983384623115578\\
400	0.994042763819095\\
500	0.99884743718593\\
750	0.993879095477387\\
1000	0.994079798994975\\
2000	0.984090452261306\\
};
\addlegendentry{$\text{LBA }\theta{}^\text{E}\text{ = 60}^\circ$}

\addplot [color=black, line width=2.0pt, mark=diamond, dashed, mark options={solid, black}]
  table[row sep=crcr]{%
1	0.994408140703518\\
5	0.999552914572864\\
10	0.994377688442211\\
25	0.994499849246231\\
50	0.994484572864322\\
100	0.994484572864322\\
150	0.989479497487437\\
200	0.989504673366835\\
250	0.988978291457286\\
400	0.993371206030151\\
500	0.993242060301507\\
750	0.983969296482412\\
1000	0.993794271356784\\
2000	0.994015326633166\\
};
\addlegendentry{$\text{CBA }\theta^\text{E}\text{ = 45}^\circ$}

\end{axis}

\end{tikzpicture}%
\vspace{-3mm}
\caption{Comparison of accuracy under code-based attack (CBA) and location-based attack (LBA), using \ac{OC-SVM}, $\theta^{\text{A}} = 0\degree$. }\vspace{-5mm}
\label{fig:acc_comparison}
\end{figure}
%The classification accuracy is plotted for both location-based and code-based attacks in Fig.~\ref{fig:acc_comparison}. 
In the context of the location-based attack, a parallel trend to the observed RMSE patterns is evident: as the distance $d^{\text{E}}$ between Eve and the verifier increases, the classification accuracy improves. This improvement occurs because Eve's ability to execute a successful impersonation attack diminishes at greater distances due to power limitations. Similarly, smaller values of $\theta^{\text{E}}$ lead to more successful attacks, resulting in reduced classification accuracy. This trend is  evident in the lower classification accuracy observed for $\theta^{\text{E}}= 5 \degree$ in comparison to that observed for $\theta^{\text{E}}= 20 \degree$, which in turn is smaller when compared to that of $\theta^{\text{E}}= 45 \degree$, aligning with the \ac{RMSE} trends illustrated in Fig. \ref{fig:estimatedAoA}. For the case of $\theta^{\text{E}}= 30 \degree$, the null of Eve's beam response aligns precisely with the main lobe of Alice's beam response, leading to the failure of impersonation attempts and consequently achieving the highest authentication accuracy.
For the code-based attack, 
%In the scenario of a code-based attack, 
the impersonation attempt is less effective, resulting in a  high classifier accuracy even with a smaller RMSE for $\theta^{\text E}$. This disparity arises because the location-based attack causes the disappearance of the original minimum and introduces a new minimum at $\theta^{\text A}$, providing unbiased estimates with a variance dependent on Eve's angle and distance. {Conversely, the code-based attack introduces bias into the estimates, making them easier to detect. }

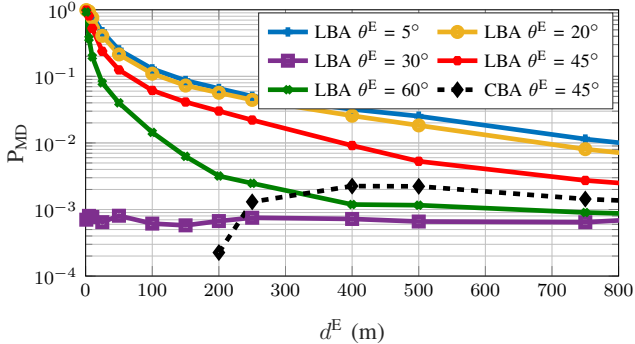
\begin{figure}%[ht]
\centering
\definecolor{mycolor1}{rgb}{0.00000,0.44700,0.74100}%
\definecolor{mycolor2}{rgb}{0.92900,0.69400,0.12500}%
\definecolor{mycolor3}{rgb}{0.49400,0.18400,0.55600}%
\definecolor{mycolor4}{rgb}{0.00000,0.49804,0.00000}%
\begin{tikzpicture}[scale=1\columnwidth/10cm,font=\footnotesize]
\begin{axis}[%
width=8 cm,
height=4 cm,
at={(0.932in,0.642in)},
scale only axis,
xmin=0,
xmax=800,
ylabel style={yshift=-0.2cm},
xlabel style={font=\color{white!15!black}},
xlabel={$d^{\text E}$ (m)},
ymode=log,
ymin=0.0001,
ymax=1,
yminorticks=true,
ylabel style={font=\color{white!15!black}},
ylabel={$\text{P}_{\text{MD}}$},
axis background/.style={fill=white},
xmajorgrids,
ymajorgrids,
yminorgrids,
legend style={legend cell align=left, align=left, legend columns=2, fill opacity=0.6, draw=white!15!black, at={(0.99,0.99)}}
]
\addplot [color=mycolor1, line width=2.0pt, mark=+, mark options={solid, mycolor1}]
  table[row sep=crcr]{%
1	0.99\\
5	0.9729522\\
10	0.8178016\\
25	0.4682832\\
50	0.2500369\\
100	0.1296801\\
150	0.08661\\
200	0.0657558\\
250	0.0502815\\
400	0.0320256\\
500	0.02523\\
750	0.0114314\\
1000	0.0061278\\
2000	0.0016709\\
};
\addlegendentry{LBA $\theta{}^\text{E}\text{ = 5}^\circ$}

\addplot [color=mycolor2, line width=2.0pt, mark=o, mark options={solid, mycolor2}]
  table[row sep=crcr]{%
1	0.99\\
5	0.9425836\\
10	0.7622277\\
25	0.4061512\\
50	0.2134\\
100	0.1097415\\
150	0.0727749\\
200	0.0566949\\
250	0.04431\\
400	0.0256113\\
500	0.0183348\\
750	0.00806000000000001\\
1000	0.0044253\\
2000	0.001247\\
};
\addlegendentry{LBA $\theta{}^\text{E}\text{ = 20}^\circ$}

\addplot [color=mycolor3, line width=2.0pt, mark=square, mark options={solid, mycolor3}]
  table[row sep=crcr]{%
1	0.000703\\
5	0.0008022\\
10	0.0007725\\
25	0.0006454\\
50	0.0008062\\
100	0.0006138\\
150	0.0005782\\
200	0.00067\\
250	0.0007524\\
400	0.0007211\\
500	0.0006596\\
750	0.0006435\\
1000	0.00088\\
2000	0.0006039\\
};
\addlegendentry{LBA $\theta{}^\text{E}\text{ = 30}^\circ$}

\addplot [color=red, line width=2.0pt, mark=asterisk, mark options={solid, red}]
  table[row sep=crcr]{%
1	0.99\\
5	0.8085882\\
10	0.5245344\\
25	0.2379507\\
50	0.1250804\\
100	0.0614656\\
150	0.0411796\\
200	0.03011\\
250	0.0220794\\
400	0.0091377\\
500	0.0053001\\
750	0.0027324\\
1000	0.00177\\
2000	0.00094\\
};
\addlegendentry{LBA $\theta{}^\text{E}\text{ = 45}^\circ$}

\addplot [color=mycolor4, line width=2.0pt, mark=x, mark options={solid, mycolor4}]
  table[row sep=crcr]{%
1	0.92414\\
5	0.3705282\\
10	0.1964367\\
25	0.08003\\
50	0.04\\
100	0.0144045\\
150	0.00629\\
200	0.0031878\\
250	0.0024735\\
400	0.001188\\
500	0.00116\\
750	0.0009009\\
1000	0.0007765\\
2000	0.0008245\\
};
\addlegendentry{LBA $\theta{}^\text{E}\text{ = 60}^\circ$}

\addplot [color=black, line width=2.0pt, mark=diamond, dashed, mark options={solid, black}]
  table[row sep=crcr]{%
1	0\\
5	0\\
10	0\\
25	0\\
50	0\\
100	0\\
150	0\\
200	0.0002254\\
250	0.0013034\\
400	0.0022473\\
500	0.0022275\\
750	0.0014356\\
1000	0.0011286\\
2000	0.0008118\\
};
\addlegendentry{CBA $\theta{}^\text{E}\text{ = 45}^\circ$}

\end{axis}

\end{tikzpicture}%
\vspace{-4mm}
\caption{Comparison of the probability of MD under code-based attack (CBA) and location-based attack (LBA), using \ac{OC-SVM} , $\theta^{\text{A}} = 0\degree$. }
\vspace{-5mm}
\label{fig:pmd_comparison}
\end{figure}

%To gain further insight,
We also analyze the probability of \ac{MD}, which is illustrated in Fig. \ref{fig:pmd_comparison} for both types of attacks. With a location-based attack, a similar trend is observed in terms of \ac{MD} probability as in accuracy: the probability of \ac{MD} for $5 \degree$  is larger than that for $20 \degree$, which in turn is larger than that for $ 45 \degree$. Notably, for $\theta^{\text{E}}= 30 \degree$, the probability of \ac{MD} is lower, aligning with the already observed accuracy trends. The probability of \ac{FA} is not illustrated in the figures, as it remains consistently around 0.014 regardless of changes in Eve's distance and \ac{AoA}.\footnote{In the specific scenario under consideration, an \ac{OCC} is trained solely on samples from Alice. Given that these samples remain unchanged from the training phase to the verification phase, except for the random noise affecting \ac{AoA} estimation, Eve's parameters have no impact on the probability of false alarm.}

%\begin{figure}[ht]
%    \centering
%    \includegraphics[width=\linewidth]{Figures/new_results_theta_minus10.eps}
%    \caption{Classification accuracy and probabilities of false alarm and missed detection with \ac{OC-SVM}, $\theta^{\text{A}} = 0$, $\theta^{\text{E}} = 10$.
%    \\
%    \hui{I believe the heatmap of accuracy within a certain area would be more meaningful, 0-5000 meters is too large, maybe change to 0-1000? Also this figure can only show one angle.}}
%    \label{fig:new_results}
%\end{figure}

%\begin{figure}[th]
%    \centering
%    \includegraphics[width=\linewidth]{Figures/new_results_zoom.eps}
%    \caption{Detail of Fig. \ref{fig:new_results}, with focus on short distances. 
%    \\
%    \hui{This figure could be replaced with heatmap of PMD, since PFA is always 0.}
%    }
%    \label{fig:zoom}
%\end{figure}

\section{Conclusions}
We studied a physical layer authentication protocol in which a \ac{BS} equipped with an analog array aims at identifying a legitimate transmitting node using an \ac{OC-SVM} classifier trained on the estimated \ac{AoA} of the same node. 
We introduced several attack techniques that could be exploited by a malicious node to forge the estimated \ac{AoA} and impersonate the legitimate node.
We studied the effectiveness of these attacks on an ML-based classifier's performance for various distances and angles, observing that location-based and code-based attacks can be successful in impersonating the \ac{AoA}. Our study reveals that a successful impersonation requires knowledge of the location of the attacker and the victim, as well as the combiners at the verifier. The effectiveness of the attack depends on the available transmission power at the attacker, as well as the nulls of the verifier's beams. {Future studies can include investigating authentication spoofing with location mismatches and exploring the impact of multipath channels.}

\bibliographystyle{IEEEtran}
\bibliography{references}

\end{document}